\documentclass[aps,pre,preprint,floatfix,showpacs,preprintnumbers,nofootinbib,eqsecnum,cernpreprint]{revtex4-1}

\pdfoutput=1

\usepackage[dvips,final]{graphicx}
\usepackage{amssymb}
\usepackage{amsmath}
\usepackage{amsfonts}
\usepackage{epsfig}
\usepackage{bm}
\usepackage{braket}
\usepackage{comment}
\usepackage{float}
\usepackage[utf8]{inputenc}
\usepackage{hhline}
\usepackage[normalem]{ulem}
\usepackage[export]{adjustbox}
\usepackage[usenames,dvipsnames,table,xcdraw]{xcolor}
\usepackage{array}
\usepackage{multirow}
\usepackage{booktabs}

\usepackage{enumerate}

\usepackage{amsmath}
\usepackage{bbold}
\usepackage{hyperref}
\hypersetup{colorlinks=true, linkcolor=teal, urlcolor=blue, citecolor=red}
\usepackage{enumerate}
\usepackage{mathtools}
\usepackage{tikz}
\usepackage[capitalise]{cleveref}

\crefname{section}{Sec.}{Secs.}
\crefname{table}{Tab.}{Tabs.}
\crefname{figure}{Fig.}{Figs.}
\crefname{equation}{Eq.}{Eqs.}
\crefname{appendix}{Appendix\ }{Appendix\ }

\definecolor{bostonuniversityred}{rgb}{0.8, 0.0, 0.0}

\definecolor{darkGreen}{RGB}{34,139,34}

\newcommand {\qaoax} 	{QAOA$_X$} 
\newcommand {\qaoaxy} 	{QAOA$_{XY}$} 
\newcommand {\qaoamis} {QAOA$_{\mathrm{MIS}}$}

\newcommand	{\Nb}		{N_b}

\newcommand	{\NHH}	{N_{\text{HH}}}
\newcommand	{\EHP}	{E_{\text{HP}}}
\newcommand	{\EHPmin}	{E_{\text{HP}}^{\text{min}}}
\newcommand	{\EXY}	{E_{XY}}
\newcommand  {\EMIS} 	{E_{\text{MIS}}} 

\newcommand  {\bv}           {\mathbf{b}}

\newcommand	{\ev}[1]	{\langle#1\rangle}
\definecolor{darkpastelgreen}{rgb}{0.01, 0.75, 0.24}

\begin{document}


\title{Penalty-free quantum optimization applied to\\ lattice protein folding}

\author{Leif Gellersen}
\email{leif.gellersen@gmail.com}
\affiliation{Department of Physics, Lund
University, SE-223 62 Lund, Sweden\\Now employed at Axis Communications}

\author{Anders Irbäck}
\email{anders.irback@mgeo.lu.se}
\affiliation{Computational Science for Health and Environment (COSHE), Lund University, Sölvegatan 12, SE-223 62 Lund, Sweden}

\author{Lucas Knuthson}
\email{lucas.knuthson@mgeo.lu.se}
\affiliation{Computational Science for Health and Environment (COSHE), Lund University, Sölvegatan 12, SE-223 62 Lund, Sweden}

\author{Stefan Prestel}
\email{stefan.prestel@quantum-brilliance.com}
\affiliation{Quantum Brilliance GmbH, Colorado Tower Industriestraße 4, DE-70565 Stuttgart, Germany}

\begin{abstract}
Identifying minimum-energy structures of lattice proteins is a challenging discrete optimization problem. 
Quantum approaches such as analog quantum annealing and the gate-based quantum approximate optimization 
algorithm (QAOA) can address this problem after mapping it to a binary representation, which 
typically involves introducing penalty terms to enforce valid chain configurations. However,
in this and many related problems,  the use of quadratic penalty terms 
can be avoided by restricting the search space to independent sets 
in a conflict graph and using a QAOA mixer designed for the maximum independent set problem. In this work,
we implement and explore this QAOA variant for lattice protein folding. Here, 
the objective function consists solely of the protein energy 
together with a simple linear bias term, without quadratic penalties.         
We validate this approach through classical simulations of the quantum circuits for lattice proteins of lengths 
$N=4$ and $N=6$. To explore larger systems, we further introduce a heuristic iterative local-search scheme, with 
which we successfully fold lattice proteins with lengths up to $N=14$ using local subgraphs with at most 26 qubits.
\end{abstract}


\maketitle

\section{Introduction}\label{sec:Intro}
Quantum computing offers a promising framework for addressing hard discrete optimization problems. 
Two leading paradigms are analog quantum annealing (QA)~\cite{Kadowaki:98} and the gate-based 
quantum approximate optimization algorithm (QAOA)~\cite{Farhi:14}. In both approaches, the 
optimization problem is encoded in a Hamiltonian that is diagonal in the computational 
basis, while a mixer Hamiltonian drives transitions between computational basis states.   
Constraints are typically enforced by incorporating penalty energies into the problem Hamiltonian. 
However, with a suitably designed mixer, the quantum evolution can in some cases be restricted 
to a subspace such that some or all penalty energies can be removed.
Within QAOA, this strategy is known as the quantum alternating operator ansatz~\cite{Hadfield:19}, 
which employs mixers beyond the standard transverse-field ($X$) mixer used in the original formulation. 
A well-known example is the $XY$ mixer~\cite{Hen:16,Hadfield:19,Wang:20}, which preserves the Hamming 
weight of the states on which it acts.    

Lattice protein folding provides a biophysically motivated example of a hard discrete optimization problem. 
The objective is to determine minimum-energy chain conformations for a given amino acid sequence.
This problem has been studied using QA~\cite{Perdomo-Ortiz:12,Irback:22}, QAOA~\cite{Fingerhuth:18,Boulebnane:23}
and variational quantum eigensolver (VQE) techniques~\cite{Robert:21,Pamidimukkala:24}, employing various binary encodings. 
A common choice is use turn-based encodings, in which the bits specify 
successive turns along the protein chain~\cite{Perdomo-Ortiz:12,Fingerhuth:18,Robert:21,Boulebnane:23,Pamidimukkala:24}. 
In this work, we instead adopt a field-like 
binary encoding~\cite{Irback:22} (see Sec.~\ref{sec:methods_qubo}), which facilitates the treatment 
of nonlocal interactions, including self-avoidance constraints. With this encoding, the complete objective function,
including all penalty terms, can be expressed in closed quadratic form. In prior work, we showed how the
resulting quadratic unconstrained binary optimization (QUBO) problem can be solved using D-Wave's quantum 
and hybrid quantum-classical annealers, for lattice proteins in two~\cite{Irback:22}
and three~\cite{Irback:25} dimensions. Methods for lattice protein folding  are also relevant for the 
inverse problem~\cite{Mulligan:20,Irback:24,Panizza:24,Linn:25}, known as protein design, where the task is to find 
sequences that fold to a given target structure.     


In this paper, we investigate a QAOA variant for lattice protein folding, which is based on a general scheme 
for restricting the search space in such a way that any quadratic penalties can be eliminated. 
This variant assumes access to a mixer that preserves the resulting feasible subspace. 
We consider QAOA rather than QA because gate-based systems offers more 
flexibility in the choice of mixer, or driver, compared to currently existing annealers. 

The QAOA variant we study is based on mapping quadratic penalty terms 
onto edges in a conflict graph, whose nodes represent the (qu)bits of the system. 
A bit configuration can be valid only if its active bits form an independent set in 
this graph; there must be no pair of active bits linked by an edge. A mixer 
preserving the subspace of such states is 
known~\cite{Hadfield:19}, and was implemented and tested for the maximum 
independent set (MIS) problem~\cite{Tomesh:22,Saleem:23}. We refer to the QAOA variant 
with this mixer as \qaoamis. When applying this variant to lattice 
protein folding, the objective function consists of the protein
energy and a linear bias term, where the latter serves to enforce the formation of   
full-length chain structures. 

We investigate the \qaoamis\ method using the two-dimensional HP lattice protein model~\cite{Lau:89,Yue:95} 
as a testbed. In this model, proteins are represented as self-avoiding chains of hydrophobic 
(H) and polar (P) residues. We focus first on two short HP sequences with lengths $N=4$ and $N=6$, 
for which classical simulation of the quantum circuits remains tractable. When increasing $N$,   
such simulations rapidly become infeasible due to high qubit and gate numbers~\cite{Linn:24}.    

To enable the study of larger systems, we formulate and implement a heuristic iterative scheme based on 
\qaoamis\ optimizations over local neighborhoods in the conflict graph. Prior work developed such a
scheme for the MIS problem~\cite{Tomesh:22}, which we adapt for lattice protein folding where the 
protein energy is to be minimized. Decomposing large problems into smaller subproblems amenable to 
quantum processing is also central to D-Wave's hybrid quantum-classical annealers~\cite{McGeoch:20b}.

We evaluate the quantum local search (QLS) scheme on HP sequences with lengths up to $N=14$, using
local neighborhoods with $\leq$26 qubits. The results are benchmarked against exact solutions 
obtained via exhaustive enumeration of all possible chain
conformations~\cite{Irback:02,Holzgrafe:11}. Although the success rate drops with 
increasing $N$, the QLS method finds the ground state for all sequences studied ($N\leq14$). 
It is thus able to provide correct solutions for problem sizes that are well beyond what can be tackled
with full-scale \qaoamis\  ($N\lesssim 6$). 






\section{Methods}\label{sec:methods}


This section begins with brief descriptions of the HP protein model,  the field-like binary representation that we use, and
the QAOA approach. We then present the \qaoamis\ variant, which uses a mixer designed for the MIS problem, and 
the QLS scheme for iterative local search with \qaoamis. The section ends with a brief summary of computational details.    

\subsection{Biophysical model}\label{sec:methods_hp}

We consider the minimal HP model for protein folding~\cite{Lau:89,Yue:95}, in which the protein is 
represented by a self-avoiding chain of hydrophobic (H) or polar (P) amino acids, or beads,
on a two-dimensional square lattice. Two beads are said to be
in contact if they are nearest neighbors on the lattice, but not along the chain. Each chain
conformation is assigned an energy defined as $\EHP=-\NHH$, where $\NHH$ is the number of
HH contacts~\cite{Lau:89}. With this definition, low-energy conformations tend
to exhibit a hydrophobic core of mainly H beads. In what is commonly referred to as 
lattice protein folding, the task is to identify the chain conformation(s) with minimum energy
for a given amino acid sequence, which has been shown to be an NP-complete problem~\cite{Crescenzi:98}.  
The ground state may or may not be unique, depending on the sequence. From exhaustive 
enumerations of all possible states, it is known that about 2\% of all HP sequences
with length $N\le$30 have a unique ground state~\cite{Irback:02,Holzgrafe:11}. 



\subsection{QUBO representation}\label{sec:methods_qubo} 

To solve the lattice protein folding problem using quantum computing, we need to map chain structures 
onto a system of classical bits. A common choice is to use turn-based representations, in which the bits 
encode the directions of the links along the chain~\cite{Perdomo-Ortiz:12,Robert:21}.  
In this paper, we instead use a field-like representation~\cite{Irback:22}, which greatly facilitates the 
implementation of non-local interactions between distant parts of the chain. Here, a bit $b_{i,n}$ 
indicates whether ($b_{i,n}=1$) or not ($b_{i,n}=0$) amino acid $i$ along the chain is located on lattice 
site $n$. 

The energy function to be minimized has the form
\begin{equation}
	E=\EHP+\lambda_1 E_1+ \lambda_2 E_2\ + \lambda_3 E_3,
	\label{eq:E}
\end{equation} 
where $\EHP$ is the HP energy (see above) and the remaining three terms $E_1$, $E_2$ and $E_3$ are penalty energies
needed to ensure that the bit configurations correspond to valid chain structures. The strengths of the penalty terms 
are set by the Lagrange parameters $\lambda_i$. All the four energy terms in Eq.~\ref{eq:E} have quadratic closed-form 
expressions valid for any chain length, which are as follows~\cite{Irback:22}.  

\begin{itemize}

\item The HP energy $\EHP=-\NHH$ can be written as
\begin{equation}
\label{eq:EHP}
\EHP=-\sum_{| i-j |>1}C(a_i,a_j)\sum_{\langle n,m\rangle}b_{i,n}b_{j,m},
\end{equation}
where $C(a_i,a_j)$ denotes the contact energy for two beads $i$ and $j$ with 
amino types $a_i$ and $a_j$, respectively. In the HP model, $C(a_i,a_j)=1$ 
if $a_i=a_j=\text{H}$ and $C(a_i,a_j)=0$ otherwise. In Eq.~\ref{eq:EHP},
$\ev{\cdot,\cdot}$ indicates summation over nearest-neighbor pairs of sites.

\item The first penalty energy, $E_1$, is given by
\begin{equation}
E_1= \sum_i\left(\sum_n b_{i,n}-1\right)^2 = \sum_i\left( \sum_{n\ne m} b_{i,n} b_{i,m} -\sum_n b_{i,n} + 1\right),
\label{eq:E1}
\end{equation}
and serves to ensure that each bead is located at exactly one lattice site. 

\item The energy $E_2$ enforces chain self-avoidance, and is given by
\begin{equation}
E_2=\sum_n\sum_{i<j} b_{i,n} b_{j,n}.
\label{eq:E2}
\end{equation}

\item The third and final penalty energy, $E_3$, is responsible for connecting the beads to a chain,
and can be written as 
\begin{equation}
E_3= \sum_{1\le i<N}\sum_n b_{i,n}
\sum_{||m-n||>1}b_{i+1,m},
\label{eq:E3}
\end{equation}
where $||m-n||$ denotes the Eucledian distance between the sites $m$ and $n$
(assuming unit lattice spacing).     
\end{itemize}

For a chain with $N$ amino acids on an $L^2$ grid, this mapping requires $\Nb=NL^2$ bits. 
With a checkerboard division of the grid into odd and even sites, it is often possible 
to reduce the bit count to $\Nb\approx$$NL^2/2$. 

The same QUBO formulation can be used for any lattice protein model with contact interactions.  
For instance, it was recently used to fold three-dimensional lattice proteins with a 20-letter amino acid alphabet~\cite{Irback:25}.

The three penalty terms represent equality constraints; a chain configuration is valid if and only if $E_1=E_2=E_3=0$.
An oracle-based penalty-free method for incorporating inequality constraints into QAOA was recently developed, and
applied to the knapsack problem~\cite{Bucher:25}.   


\subsection{QAOA}\label{sec:methods_qaoa}

We wish to determine minimum-energy structures of HP proteins using QAOA and 
the binary representation in Sec.~\ref{sec:methods_qubo}. 
The state of the binary system is specified by a string $\bv$ 
of bits $b_{i,n}$, whose 0 and 1 states correspond qubit states $\ket{b_{i,n}}$  
defined by $\frac{1}{2}(1-Z_{i,n}) \ket{b_{i,n}}=b_{i,n} \ket{b_{i,n}}$ [throughout the paper, 
$X_{i,n}, Y_{i,n}$ and $Z_{i,n}$ denote Pauli operators acting on qubit ($i,n$)]. 
The product states $\ket{\bv}=\prod_{i,n}\ket{b_{i,n}}$ are referred to as the computational basis. 


QAOA in its original form~\cite{Farhi:14} performs an unconstrained optimization of an 
objective function $E(\bv)$ (in our case given by Eq.~\ref{eq:E}) over all possible bitstrings $\bv$. 
To this end, one introduces a cost Hamiltonian $H$, which is diagonal in the computational basis 
and defined by $H\ket{\bv}=E(\bv)\ket{\bv}$, and a mixer Hamiltonian $M_X=-\sum_{i,n} X_{i,n}$, 
which generates transitions between the computational basis states. Starting from a given initial state 
$\ket{\psi_0}$, typically a uniform superposition of all the computational basis states, 
the qubit system evolves under an alternating sequence of the unitary operators 
$U_H(\gamma)=e^{-i\gamma H}$  and $U_M(\beta)=e^{-i\beta M_X}$. The resulting final state $\ket{\psi_p}$ 
is given by 
\begin{equation}
\ket{\psi_p}=\ket{\psi_p(\boldsymbol{\theta})}=U_M(\beta_p)U_H(\gamma_p)\ldots U_M(\beta_1)U_H(\gamma_1)\ket{\psi_0},
\label{eq:evo}\end{equation}
where the $2p$ angles $\boldsymbol{\theta}=(\beta_1,\ldots,\beta_p, \gamma_1,\ldots,\gamma_p)$ 
are variational parameters. This is followed by measurement  in the computational basis
and repeated many times for a given $\boldsymbol{\theta}$.
The measured bitstrings are used to estimate the expectation of $H$ in 
the state $\ket{\psi_p(\boldsymbol{\theta})}$, 
\begin{equation}
F(\boldsymbol{\theta})=\braket{\psi_p(\boldsymbol{\theta})| H |\psi_p(\boldsymbol{\theta})}.
\label{eq:F}
\end{equation} 
By iteratively minimizing $F(\boldsymbol{\theta})$, an approximate
minimum-$E$ solution is generated~\cite{Farhi:14}.

In some cases, it is possible to restrict the quantum evolution (Eq.~\ref{eq:evo}) to a subspace   
where one or more constraints are fulfilled~\cite{Hadfield:19}, to which the initial state 
$\ket{\psi_0}$ is assumed to belong. This requires that the mixer $U_M(\beta)$ preserves this subspace. 
A well-known example is the $XY$-mixer~\cite{Hen:16,Hadfield:19,Wang:20}, 
which preserves the Hamming weight of the states on which it acts. 
In particular, this mixer can be used to implement one-hot constraints like those softly 
enforced by the penalty term $E_1$ in Eq.~\ref{eq:E1}.
Specifically, if we set        
\begin{equation}
\label{eq:m_XY}
U_M(\beta) = \prod_i\prod_{n<m} e^{i\beta(X_{i,n}X_{i,m} +Y_{i,n}Y_{i,m})}.
\end{equation}      
and use a properly chosen $\ket{\psi_0}$, then the objective function can simplified to
\begin{equation}
\EXY=\EHP+ \lambda_2 E_2\ + \lambda_3 E_3.
\label{eq:EXY}
\end{equation}      
We define the initial state $\ket{\psi_0}$ as a tensor product of one-bead states, 
where each bead is in a uniform superposition of all its feasible states (those 
with Hamming weight one). This superposition is a simple example of a Dicke state~\cite{Bartschi:19}. 
Note that the two-local factors in Eq.~\ref{eq:m_XY} do not commute. In our computations, we use a fixed random 
ordering of these factors for each problem instance.    

We will refer to standard QAOA with its $X$-mixer and the variant using the $XY$-mixer 
as \qaoax\ and \qaoaxy, respectively.  

In optimizing the variational parameters $\boldsymbol{\theta}=(\boldsymbol{\beta},\boldsymbol{\gamma})$,
we follow the same procedure for both  \qaoax\ and \qaoaxy.
We begin with a grid-based exploration of the two-dimensional cost landscape $F(\beta,\gamma)$ (Eq.~\ref{eq:F})
for a single-layer circuit ($p=1$). Using low-cost points from this scan as initial values, the $p=1$ circuit is 
optimized. We then iteratively optimize circuits with increasing depth $p$ using an interpolation-based scheme 
for parameter transfer from smaller to larger circuits~\cite{Zhou:20}. 
 
 
 \subsection{\qaoamis}\label{sec:methods_qaoamis}
 
\qaoaxy\ works with a reduced search space such 
that the penalty term $E_1$ can be removed. In this subsection, we describe how the 
search space can be further reduced by mapping the problem onto a conflict graph and using 
a mixer designed for the MIS problem~\cite{Hadfield:19,Tomesh:22}.    
The resulting QAOA variant is referred to as \qaoamis. 
 
\begin{figure}
\centering
   \includegraphics[width=3.5cm,valign=c]{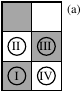}
   \hspace{1cm}
   \includegraphics[width=6.5cm,valign=c]{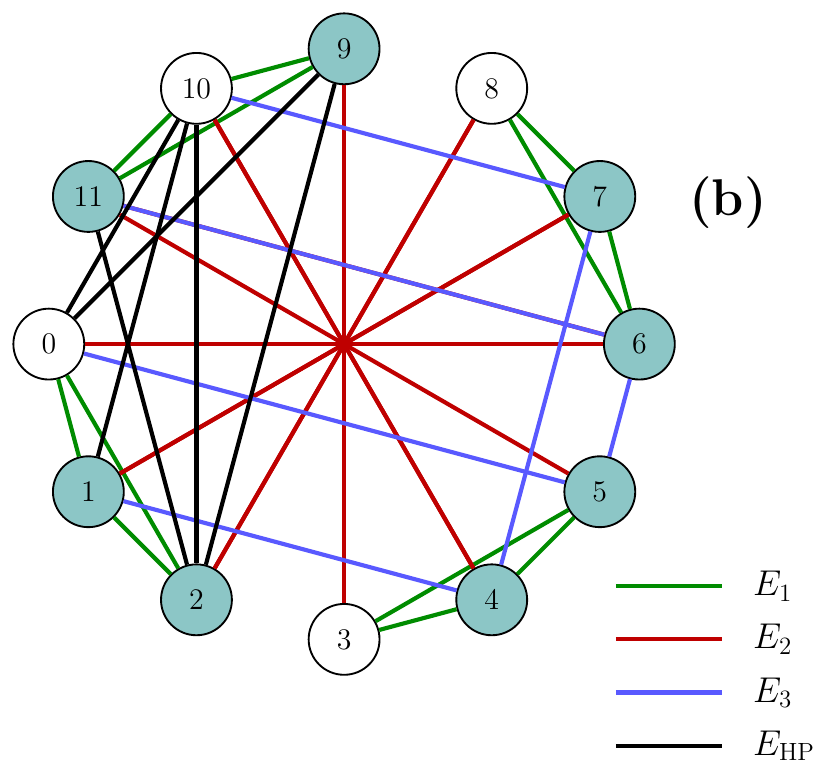}
    \caption{Folding the sequence HPPH on a $3\times2$ lattice. (a) A valid chain conformation with minimum energy 
    ($\EHP=-1$). The amino acids are numbered I--IV.  With a checkerboard division of the sites, odd and 
    even amino acids can be assumed to reside on sites with different color. The QUBO representation in 
    Sec.~\ref{sec:methods_qubo} then requires three bits per amino acid, that is 12 bits in total.  
    (b) Graph with the 12 bits at the nodes and with edges corresponding to quadratic energy terms (Eq.~\ref{eq:E}).
    Black edges represent favorable interactions, coming from $\EHP$.  The other edges represent conflicts 
    stemming from the penalty energies $E_1$ (green), $E_2$ (red) and $E_3$ (blue). Node color indicates 
    whether the bit is 1 (white) or 0 (green) in the state shown in (a). 
    \label{fig:qubo}}
\end{figure}

The conflict graph is obtained by associating the bits  
$b_{i,n}$ with nodes and indicating all quadratic terms in  
$E_1$ (Eq.~\ref{eq:E1}), $E_2$ (Eq.~\ref{eq:E2}) and $E_3$ (Eq.~\ref{eq:E3}) by  
edges (Fig.~\ref{fig:qubo}). 
A bit configuration corresponds to a valid chain structure if and only if all the quadratic
terms vanish and $\sum_{i,n}b_{i,n}=N$. For the former condition to be fulfilled, 
there must be no pair of active nodes (both 1) connected by an edge, which 
means that the active nodes must form an independent set. Due to the $E_1$ term, the 
maximum size of an independent set is $N$. It follows that full-length chain structures 
correspond to maximum independent sets, and that the folding problem can be 
solved by minimizing 
\begin{equation}
\EMIS=\EHP+\lambda_1 \left(N-\sum_{i,n} b_{i,n}\right)
\label{eq:EMIS}
\end{equation}
over all independent sets.

A QAOA mixer that preserves the subspace corresponding to independent sets is available~\cite{Hadfield:19},  
and was applied to the MIS problem~\cite{Tomesh:22,Saleem:23}. This mixer, $U_M(\beta)$, is given by 
\begin{equation}
U_M(\beta) = \prod_{i,n} V_{i,n}(\beta)\qquad 
V_{i,n}(\beta) = e^{-i\beta X_{i,n} \bar{B}_{i,n}}\qquad
\bar{B}_{i,n}=\prod_{(j,m)\in N_{i,n}}\frac{1+Z_{j,m}}{2},
\label{eq:mis-mix}
\end{equation}
where $N_{i,n}$ denotes the set of neighbors to qubit $(i,n)$ in the conflict graph.  
The partial mixer $V_{i,n}$ in Eq.~\ref{eq:mis-mix} can be expressed as 
\begin{equation}
\label{eq:V_alpha}
V_{i,n}(\beta)=I+(e^{-i\beta X_{i,n}}-I)\bar B_{i,n},
\end{equation} 
$I$ being the unit operator. It thus amounts to a multi-qubit controlled $X$-rotation of qubit $(i,n)$,
which is left unchanged whenever any of its neighbors is in the active $\ket{1}$ state. 
It follows that $U_M(\beta)$ indeed preserves the subspace corresponding to independent sets. 

\begin{figure}
\centering
   \includegraphics[width=6cm]{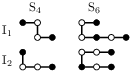}
    \caption{ Full-length chain structures corresponding to states labeled I$_1$ and I$_2$ that served as 
    initial state $\ket{\psi_0}$ (Eq.~\ref{eq:evo}) in \qaoamis\ for the sequences 
    S$_4$ and S$_6$ (Appendix~\ref{app:hp}). The Hamming distance to the ground state is minimal (2) for 
    the I$_1$ states and larger for the I$_2$ states (4 and 6, respectively). Filled and open symbols 
    indicate H and P beads, respectively.   
    \label{fig:initial_states_mis}}
\end{figure}

In implementing QAOA with this mixer for lattice proteins, we made the following choices: 
\begin{itemize}
\item \textit{Ordering of the} $V_{i,n}$'\textit{s}. The action of $U_M(\beta)$ depends on the order of the partial mixers (Eq.~\ref{eq:mis-mix}),
which do not commute. For each problem instance, we picked a fixed random ordering of the $V_{i,n}$'s.
\item  \textit{Node-dependent} $\beta$. For increased expressivity, we made the angle $\beta$ node-dependent, 
thus replacing $\beta$ with $\beta_{i,n}$ in Eq.~\ref{eq:mis-mix}. This increases the number
of variational parameters in a $p$-layer circuit from $2p$ to $(\Nb+1)p$, where $\Nb$ is the qubit count. 
For the MIS problem, it was shown that having a single layer with node-dependent $\beta_{i,n}$ can be  
beneficial compared to having more layers and one $\beta$ per layer, keeping the total number 
of variational parameters unchanged~\cite{Saleem:23}.     
\item \textit{Choice of the initial state} $\ket{\psi_0}$ \textit{(Eq.~\ref{eq:evo})}. In our $N=4$ and $N=6$ computations, we 
consider three choices of $\ket{\psi_0}$ referred to as I$_0$, I$_1$ and I$_2$, all of which 
are computational basis states in the feasible subspace. The state I$_0$ corresponds to the all-zero 
bitstring, that is no beads present on the lattice. The states I$_1$ and I$_2$ correspond to
full-length chain structures and are illustrated in Fig.~\ref{fig:initial_states_mis}. 
\end{itemize}

Table~\ref{tab:variants} provides a brief summary of the three QAOA variants considered in this paper.  
Unfortunately, implementing these methods for lattice protein folding on 
current classical or quantum hardware is possible only for short chains,  
mainly due to a rapidly growing gate count~\cite{Linn:24}. To mitigate this 
limitation, we formulate and explore an iterative heuristic scheme based on 
local searches with \qaoamis, as described in the next subsection.  

\begin{table}[t]
    \caption{\label{tab:variants} Mixer operators $U_M(\beta)$ and objective functions for the 
    three QAOA variants studied.}
\begin{ruledtabular}
\begin{tabular}{lll}
Method & Mixer $U_M(\beta)$ & Objective function \\
\hline
\qaoax  	& $\prod_{i,n}e^{i\beta X_{i,n}}$ 								& $E=\EHP+\lambda_1E_1+\lambda_2E_2+\lambda_3E_3$			\\
\qaoaxy  	& $\prod_i\prod_{n<m}e^{i\beta(X_{i,n}X_{i,m}+Y_{i,n}Y_{i,m})}$		& $\EXY=\EHP+\lambda_2E_2+\lambda_3E_3$  					\\
\qaoamis 	& $\prod_{i,n} e^{-i\beta X_{i,n}\bar B_{i,n}}$\quad 
                    $\bar B_{i,n}=\prod_{\text{nbrs}}\frac{1}{2}(1+Z_{j,m})$				& $\EMIS=\EHP-\lambda_1\sum_{i,n} b_{i,n}\ (+\mathrm{const.})$ 	\\
\end{tabular}
\end{ruledtabular}
\end{table}


\subsection{Folding longer chains using QLS}\label{sec:methods_qls}

Let $C$ be the set of nodes in the conflict graph  for a lattice protein 
whose energy we wish to minimize (Sec.~\ref{sec:methods_qaoamis}). In this subsection, 
we describe a heuristic scheme, QLS, for this task, which is based on iterative local 
search with \qaoamis\ over neighborhoods $C_{loc}\subseteq C$. 
The local search over a given $C_{loc}$ aims at minimizing $\EMIS$ (Eq.~\ref{eq:EMIS}) for fixed 
values of the bits not in $C_{loc}$. 

The state of the system is stored in a global bitstring 
$S=(\ldots,b_{i,n},\ldots)$, which may initially be set to the all-zero state 
$S=(0,\ldots,0)$. Each local search is initialized from the current $S$. 
After the search, the new values of the bits in $C_{loc}$ are 
written back to $S$. 

The action of a partial mixer $V_{i,n}$ (Eq.~\ref{eq:mis-mix}) for a 
node $(i,n)$ in a given $C_{loc}$ depends on all neighboring qubits, including frozen
ones not in $C_{loc}$. If $(i,n)$ has any frozen neighbor in the $\ket{1}$ state, 
it is itself effectively frozen in the $\ket{0}$ state, to avoid conflict. Now, if frozen in 
the $\ket{0}$ state, it has no impact on the action of other partial 
mixers. The node can therefore be removed from $C_{loc}$ without 
changing the outcome of the local search. 
  
Prior work introduced a quantum local search scheme for the MIS problem, 
which was successfully tested on graphs with up to 100 nodes~\cite{Tomesh:22}.  
Lattice protein folding differs from the MIS problem by the presence of the 
protein energy, which makes the systems prone to get trapped in 
local minima. To mitigate this issue, we developed a problem-specific procedure for 
selecting the local neighborhoods $C_{loc}$, which involves ``pinning'' and ``unpinning'' of 
amino acid. Nodes associated with pinned amino acids are excluded from the 
local search. The pinning status is dynamic and stored in a list. Initially, 
all amino acids are unpinned.  

Specifically, each local search proceeds in three steps, which are as follows.    

\begin{enumerate}

\item Selection of $C_{loc}$. We first randomly select a root node, $(i_0,n_0)$.
We then add every one of its neighbors that can be included without extending 
the search to states that are unfeasible due to conflicts with surrounding nodes (see above).
We restrict ourselves to first neighbors to keep the quantum circuit computationally manageable.  
An illustration of a local neighborhood can be found in Fig.~\ref{fig:MISandQLS}. 
The amino acid index $i_0$ is drawn from the uniform distribution on $\{1,\ldots,N\}$. 
For a chosen $i_0$, we order all lattice points, $n_0$, by neighborhood size (qubit count) 
from smaller to bigger and indexed by $k$. The index is sampled from the 
geometric distribution $P(k)=(1-p)^{k-1}p$ with $p=0.5$ to favor
smaller neighborhoods. 

\begin{figure}
\centering
   \includegraphics[width=12cm]{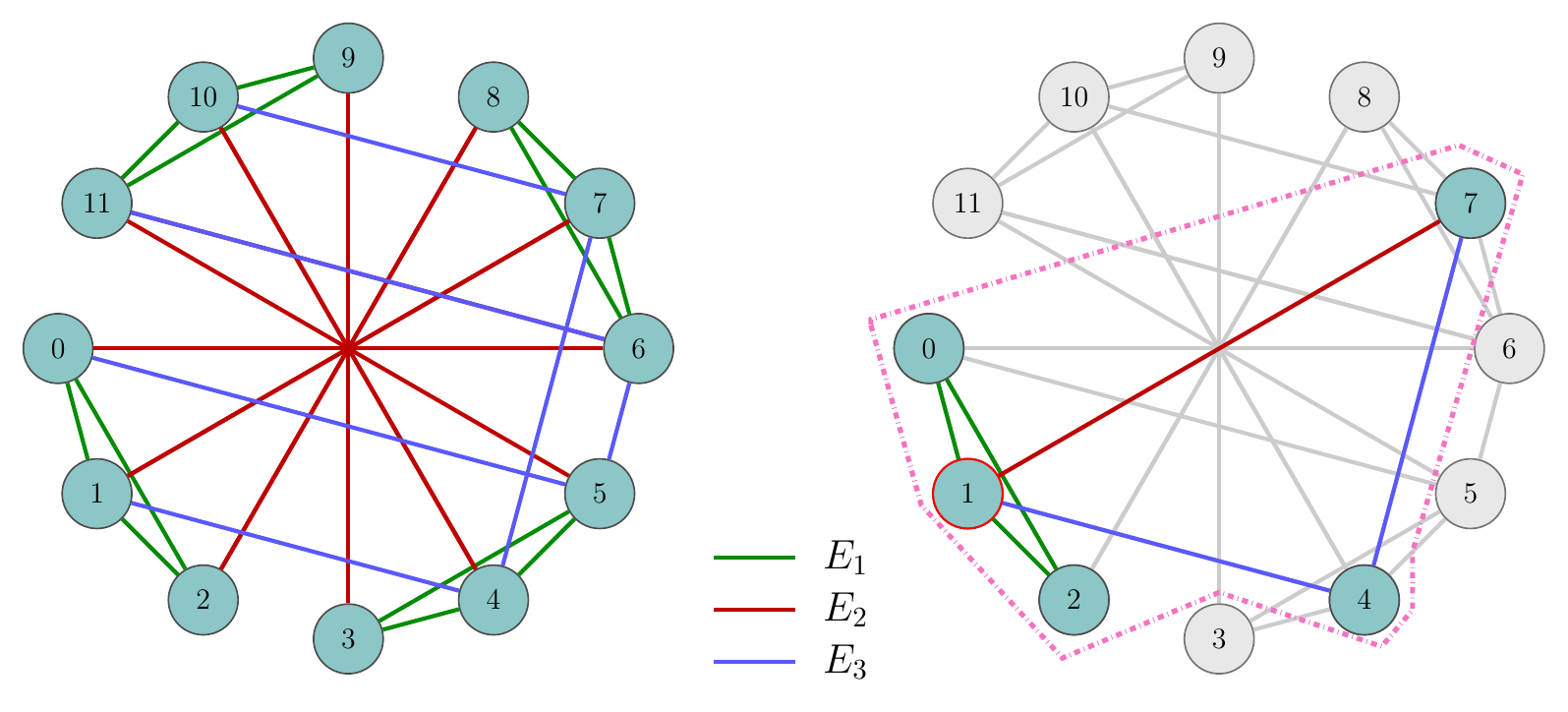}
    \caption{
         (Left) Conflict graph for the sequence HPPH on a $3\times2$ lattice. 
	The edges represent the quadratic terms in $E_1$, $E_2$ and $E_3$ (Sec.~\ref{sec:methods_qubo}). 
         The color indicates what type of conflict it is. 
         (Right) Example of a local neighborhood 
         in this conflict graph (enclosed by purple dots/dashes). It contains a root node (red circle)
         and its neighbors. Grey nodes and edges do not belong to the neighborhood.
         \label{fig:MISandQLS}}
\end{figure}

\item Pinning/unpinning of amino acids. The pinning status is updated based on the outcome of the 
preceding local search. Any amino acid $i$ for which $b_{i,n}$ changed from 0 to 1 for some $n$ 
becomes pinned with probability $p_p$. Any amino acid that was already pinned is unpinned 
with probability $p_u$. In the computations presented below, these parameters are set to $p_p=0.25$ 
and $p_u=0.10$.     

\item \qaoamis-based minimization of $\EMIS$ over unpinned amino acids in $C_{loc}$.
The initial state $\ket{\psi_0}$ (Eq.~\ref{eq:evo}) is 
chosen to be a product of single-qubit states, $\ket{\psi_{i,n}}$. We set $\ket{\psi_{i,n}}=\ket{0}$ if $b_{i,n}=0$. 
If $b_{i,n}=1$, we set $\ket{\psi_{i,n}}=2^{-1/2}\left(\ket{0}+\ket{1}\right)$ rather than $\ket{\psi_{i,n}}=\ket{1}$,  
to facilitate escape from local minima. 

\end{enumerate}


When testing this scheme for \qaoamis-based iterative local search, with its pinning mechanism, 
we first tried starting from the all-zero initial state $S=(0,\ldots,0)$, 
which, however, gave unsatisfactory results. In order to obtain a better initial state, we 
introduced a preparatory computation focusing on the placement of the H amino acids, 
which dominate the core of low-energy structures. 

The preparatory run uses \qaoamis, without local search, to pack all H segments of the full chain 
into a low-energy configuration. To obtain a suitable starting point for the computations for the 
entire chain, we added a term to the QAOA cost function $F(\boldsymbol{\theta})$
(Sec.~\ref{sec:methods_qaoa}), which provides an energy penalty if the distance between
two consecutive H segments along the chain is incompatible with the length 
of the connecting P segment. In the preparatory run, the grid consists of all sites within a 
Manhattan distance $d$ from some given site near the center of the full grid, where we set $d=2$.

\subsection{Computational details\label{sec:methods_computational_details}}
In QAOA, the aim is to minimize the expectation value 
$F(\boldsymbol{\theta})=\braket{ \psi_p(\boldsymbol{\theta})| H |\psi_p(\boldsymbol{\theta})}$ over the variational parameters 
$\boldsymbol{\theta}=(\boldsymbol{\beta},\boldsymbol{\gamma})$. We did this optimization using the Qiskit~\cite{qiskit:24} quantum software 
package to generate $\ket{\psi_p(\boldsymbol{\theta})}$ (Eq.~\ref{eq:evo}) given $\boldsymbol{\theta}$, 
and \texttt{COBYLA}~\cite{SciPy:20} to optimize $\boldsymbol{\theta}$. 

In Sec.~\ref{sec:results_short_chains} below, we test the \qaoamis, \qaoax\ and \qaoaxy\ methods on two short sequences, using 
circuits with $1\le p\le 20$ layers. In these calculations, having generated $\ket{\psi_p(\boldsymbol{\theta})}$
for a given $\boldsymbol{\theta}$, we used $10^5$ shots to estimate $F(\boldsymbol{\theta})$. In \texttt{COBYLA}, the maximum number of iterations
was set to 1,000 and the tolerance to $10^{-2}$. 

In Sec.~\ref{sec:results_longer_chains}, we test the method based on iterative local search with \qaoamis\ (Sec.~\ref{sec:methods_qls}) 
on some longer chains. In preparatory runs with H amino acids only, we used $p=3$, $10^4$ shots to estimate $F(\boldsymbol{\theta})$, 
and maximum 10,000 iterations with a tolerance of $10^{-4}$ in \texttt{COBYLA}. In the production runs, we used $p=3$, 1,000 shots to estimate 
$F(\boldsymbol{\theta})$, and maximum 1,000 iterations with a tolerance of $10^{-2}$ in \texttt{COBYLA}.


\section{Results}

We explore the utility of the graph-based \qaoamis\ method (Sec.~\ref{sec:methods_qaoamis}) for folding lattice proteins 
using classical simulations of the quantum circuits. We consider a set of HP sequences S$_N$ with lengths $4\le N\le 18$, 
which have known~\cite{Irback:02,Holzgrafe:11} unique ground states (Appendix~\ref{app:hp}). 
First, in Sec.~\ref{sec:results_short_chains}, we test full-scale \qaoamis\ on two short sequences, S$_4$ and S$_6$, 
and compare the performance to that of \qaoax\ and \qaoaxy\ (Sec.~\ref{sec:methods_qaoa}). In Sec.~\ref{sec:results_longer_chains},
we present results obtained with the heuristic QLS scheme  (Sec.~\ref{sec:methods_qls}) for sequences with $4\le N\le 18$.  
Finally, in Sec.~\ref{sec:results_resources}, we discuss resource requirements for QLS.         



\subsection{Short sequences, S$_4$ and S$_6$\label{sec:results_short_chains}}

We study and compare the depth-dependent performances of the three QAOA variants by computations for 
the sequences S$_4$ and S$_6$ on $3\times2$ and $4\times2$ grids, respectively. 

\begin{figure}[t]
\centering
	\includegraphics[width=8cm]{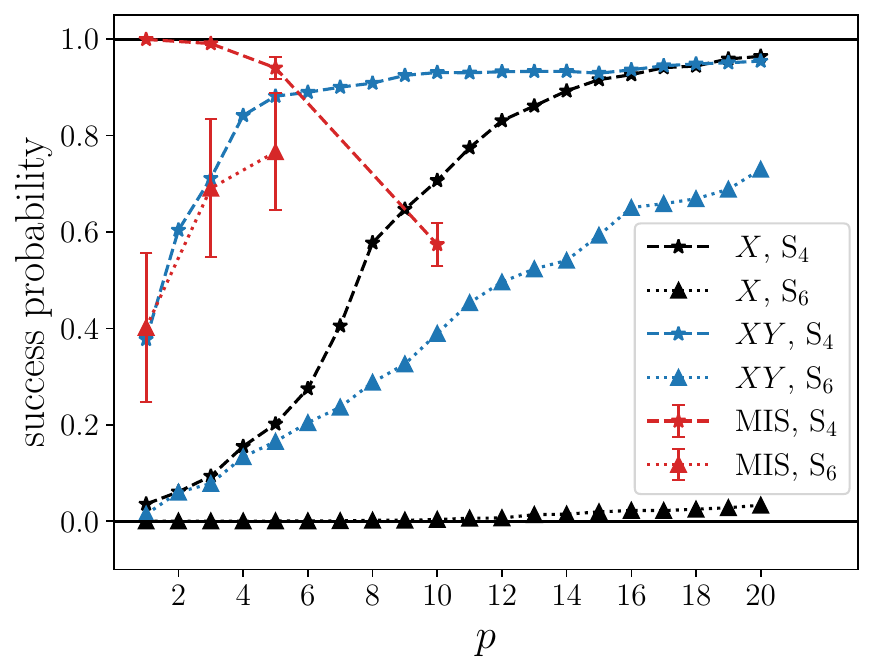}
   \caption{Success probability, defined as the squared overlap with the known ground state, as a function of 
   the circuit depth $p$ in \qaoax, \qaoaxy\ and \qaoamis\ (Table~\ref{tab:variants}) 
   simulations for the sequences S$_4$ and S$_6$ (Appendix~\ref{app:hp}) on $3\times2$ and $4\times2$ grids, respectively. 
   \qaoamis\ data are averages over 10 runs, with different random orders of the partial 
   mixers $V_{i,n}$ and different random initial values for the variational parameters
   $\boldsymbol{\beta} \in [0,2\pi]^{p\Nb}$ and $\boldsymbol{\gamma}\in [0,2\pi]^p$. 
   The initial state $\ket{\psi_0}$ was the all-zero state, I$_0$. 
   The bias was set to $\lambda_1=1$ (Table~\ref{tab:variants}).
   Error bars are standard errors of the mean. 
   For \qaoax\ and \qaoaxy, a single run was performed for each $p$. 
   Initial values for the variational parameters were obtained by a grid search for $p=1$ and by 
   parameter transfer~\cite{Zhou:20} for $p>1$ (Sec.~\ref{sec:methods_qaoa}).  
   All Lagrange parameters were set to $\lambda_i=2$ (Table~\ref{tab:variants}).   
   Lines are drawn to guide the eye. The horizontal lines indicate 0 and 1. 
   \label{fig:small_chains_hitrates}
    }
\end{figure}

We begin with the \qaoax\ and \qaoaxy\ computations. For each variant, we identified four 
initial points for the parameter optimization for depth $p=1$ through a grid-based scan of 
in the two-dimensional cost landscape $F(\beta,\gamma)$ (Eq.~\ref{eq:F}; Appendix~\ref{app:energy_landscapes}). 
For each initial point, we then iteratively
optimized circuits with increasingly large $p$ using parameter transfer by interpolation~\cite{Zhou:20},
which is a deterministic procedure. Figure~\ref{fig:small_chains_hitrates} shows results from the run 
among these four that gave the highest success probabilities. 
As expected, the success probability 
increases essentially monotonically with $p$ for both \qaoax\ and \qaoaxy.  At small $p$, 
the success probability is higher with \qaoaxy\ than with \qaoax. However, at the largest $p=20$, the 
success probability is similar ($\approx$0.95) with both methods for S$_4$. For S$_6$, 
\qaoaxy\ performs better than \qaoax\ over the whole range of $p$ studied, with success probabilities of
0.73 and 0.03, respectively, for $p=20$.  
In additional \qaoax\ computations for S$_4$, we used Trotterized annealing with 
a linear schedule rather than optimized values for the parameters ($\boldsymbol{\beta}, \boldsymbol{\gamma}$).  
With this method, we estimate that  $p\sim 10^4$ layers would be needed to achieve success probabilities $\ge$0.95,
thus indicating that parameter optimization is indeed useful. 

Figure~\ref{fig:small_chains_hitrates} also shows results obtained with \qaoamis. 
Here, the parameter optimization was randomly initialized for each $p$, with angles uniformly drawn 
on $[0,2\pi]$. The results presented are averages over 10 runs with different random number seeds,
obtained with the all-zero state (I$_0$) as the initial state $\ket{\psi_0}$ (Eq.~\ref{eq:evo}).  

The \qaoamis\ results show a very different behavior, compared to the \qaoax\ and \qaoaxy\ data 
(Fig.~\ref{fig:small_chains_hitrates}). For S$_4$, the success probability is close to one with only a single layer $(p=1)$,
but decreases when adding more layers and is just below 0.6 at the largest $p$ studied, $p=10$.   
We attribute this drop to incomplete parameter optimization. Finding the optimum starting from 
random initial points likely gets harder with increasing dimensionality of the parameter space 
[$(N_b+1)p$, $N_b$ being the qubit count].  

For S$_6$, the \qaoamis\ success probability is well below one for $p=1$, indicating that 
a single layer is insufficient to properly explore the state space of this larger system. 
Consistent with this, the success probability rises to $\approx$0.77
at the largest $p$ studied, $p=5$. However, for larger $p$, we expect the same falling trend
as for S$_4$, due to parameter optimization problems.    
Further inspection shows that the success probability of individual \qaoamis\ runs for S$_6$
tends to be close to either zero or one and thus have a bimodal distribution,  
as reflected in the large error bars. As $p$ is increased from 1 to 5, the peak near one grows 
in size, indicating improved state space exploration. 

The above \qaoamis\ results were obtained with the initial state $\ket{\psi_0}$ (Eq.~\ref{eq:evo})
chosen as the all-zero state I$_0$, corresponding to no beads present. For both S$_4$ and S$_6$, 
we also tested two choices, I$_1$ and I$_2$, corresponding to full-length chain structures
(Fig.~\ref{fig:initial_states_mis}). In these states, the bias term (Eq.~\ref{eq:EMIS}) vanishes so 
$\EMIS=\EHP\le 0$, while $\EMIS=\lambda_1N$ in the I$_0$ state. 

\begin{figure}
\includegraphics[width=8.cm]{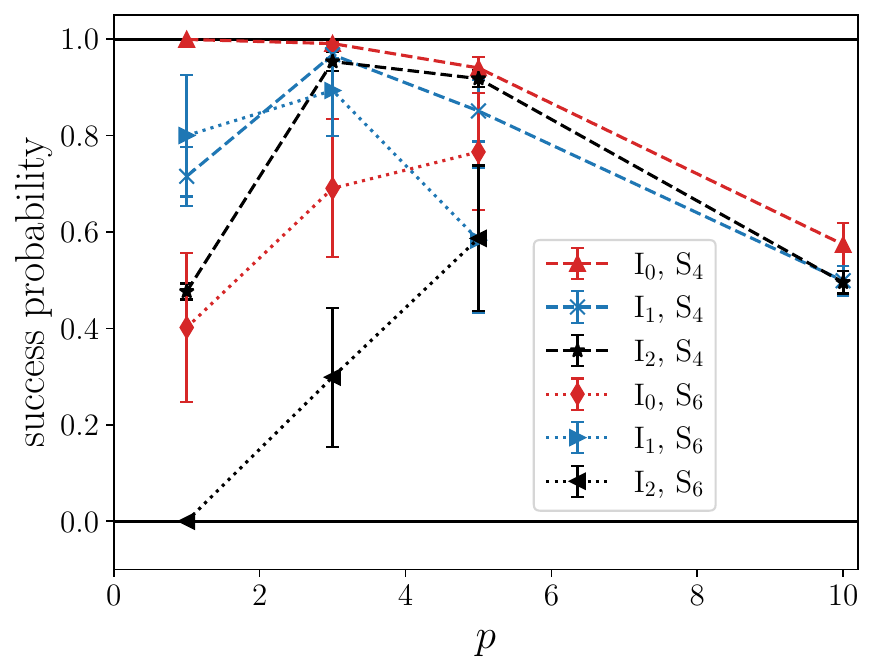} 
\caption{Success probability, defined as the squared overlap with the known ground state, as a function of 
   the circuit depth $p$ in \qaoamis\  simulations (bias $\lambda_1=1$) for the sequences S$_4$ and S$_6$ 
   (Appendix~\ref{app:hp}) on $3\times2$ and $4\times2$ grids, respectively,
   for three choices of the initial state $\ket{\psi_0}$: I$_0$, I$_1$ and I$_2$ (Fig.~\ref{fig:initial_states_mis}).  
   The data points represent averages over 10 runs, which used different random orders of the partial 
   mixers $V_{i,n}$ and were started with different random initial values for the variational parameters
   $\boldsymbol{\beta} \in [0,2\pi]^{pN}$ and $\boldsymbol{\gamma}\in [0,2\pi]^p$. Error bars show standard errors
   on the mean.  
   Lines are drawn to guide the eye. The horizontal lines indicate 0 and 1.
   \label{fig:psi_zero} 
}
\end{figure}

Figure~\ref{fig:psi_zero} compares the success probabilities obtained  
with the different initial states. While the dependence on the initial state gets weaker with 
increasing $p$, it is strong for small $p$. For $p\le 3$ and both sequences, the success probability is
higher for the I$_1$ initial state than for I$_2$, as one might expect as I$_1$ is closer than I$_2$
to the ground state (Fig.~\ref{fig:initial_states_mis}). However, for S$_4$, the best result
is for the I$_0$ initial state, which has a larger Hamming distance than I$_1$ to the ground state. 
A possible explanation for this is that there is a downhill path, in $\EMIS$, from I$_0$ to the ground state, which there  
is not for the chain structures I$_1$ and I$_2$. Such a downhill path from I$_0$ exists for S$_6$ 
as well, but is longer in this case. For S$_6$, the success probability is highest for I$_1$ with lowest 
Hamming distance (2) to the ground state. 
      

\subsection{Longer chains with QLS}
\label{sec:results_longer_chains}

Classical simulations of full-scale QAOA quantum circuits are limited to short chains, since the 
state space grows exponentially with the qubit count which in turn scales as $N^2$ or faster with the chain length $N$.      
The heuristic QLS approach described in Sec.~\ref{sec:methods_qls} aims at 
extending the range of system sizes amenable to study. To explore this approach, we conducted 
computations for chain lengths up to $N=14$, using the sequences S$_4$--S$_{14}$ 
(Appendix~\ref{app:hp}). For each sequence, we performed 10 QLS runs. In a QLS run,
the local neighborhoods are stochastically generated and vary in size. To keep the simulations 
manageable, we restricted ourselves to neighborhoods with $\le$26 qubits. When coming 
a larger neighborhood, we skipped it and generated a new one. 

\begin{figure}[t]
\centering
\includegraphics[width=0.5\textwidth]{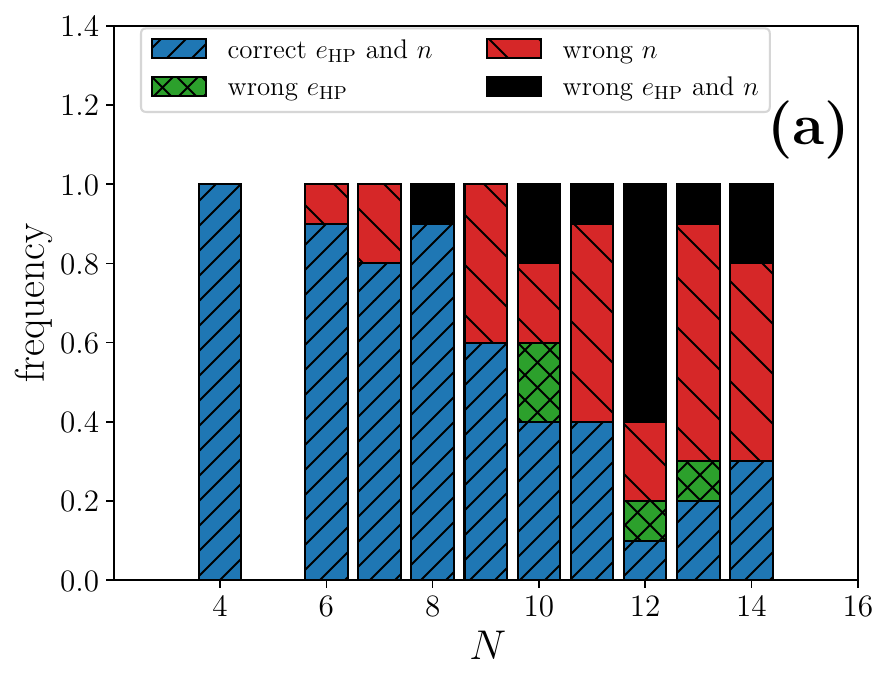}\qquad
\begin{tabular}{llcc}
     \includegraphics[width=0.35\textwidth]{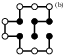} 
     \vspace{5.7 cm}
\end{tabular}
\vspace{-5.2 cm}

\caption{(a) Fractions of correct and incorrect solutions in runs with the QLS approach described in
Sec.~\ref{sec:methods_qls} for the sequences S$_4$--S$_{14}$ (Appendix~\ref{app:hp}).
For each S$_N$, 10 runs were performed. The solution generated by a run is correct 
if the number of beads, $n$, and the protein energy, $e_\mathrm{HP}$, satisfy $n=N$ and $e_\mathrm{HP}=\EHPmin$.
Incorrect solutions are divided into three groups: (i) $n=N$, $e_\mathrm{HP}>\EHPmin$, 
(ii) $n<N$, $e_\mathrm{HP}=\EHPmin$, and  (iii) $n<N$, $e_\mathrm{HP}>\EHPmin$. 
(b) Minimum-energy structure for the longest sequence studied, S$_{14}$.  
Filled and open symbols indicate H and P beads, respectively.
\label{fig:successrate_qls}}
\end{figure}

Let $n$ and $e_\mathrm{HP}$ be the number of beads and the HP energy, respectively, for
a configuration generated by a QLS run for sequence S$_N$. This solution is correct if $n=N$ 
and $e_\mathrm{HP}=\EHPmin$, where $\EHPmin$ is the known minimum energy 
of S$_N$ (Appendix~\ref{app:hp}).    
Figure~\ref{fig:successrate_qls}a 
shows the fractions of correct and incorrect solutions for S$_4$--S$_{14}$. The fraction of 
correct solutions is $\ge$0.8 for $N\le8$, and $\ge$0.1 for all sequences studied. 
The incorrect solutions can be split into three groups: 
(i) $n=N$, $e_\mathrm{HP}>\EHPmin$, (ii) $n<N$, $e_\mathrm{HP}=\EHPmin$, and 
(iii) $n<N$, $e_\mathrm{HP}>\EHPmin$. Case (ii) is the most prevalent type of error (Fig.~\ref{fig:successrate_qls}a). 
These errors can arise due to skipping larger neighborhoods, which often appear
when stitching the chain together after the formation of an H core (see Sec.~\ref{sec:methods_qls}). 

The sequence with lowest success rate is S$_{12}$, for which only 
1 of 10 runs found the ground state. Instead, a majority of the runs returned configurations 10.1007/BF01994876
with both $n < N$ and $e_\mathrm{HP} > \EHPmin$. Notably, S$_{12}$ has the largest 
average neighborhood size (see Sec.~\ref{sec:results_resources} below), which suggests that 
excluding neighborhoods with $>$26 qubits could significantly impact the success rate. 
In addition, we performed QLS computations for the sequences S$_{15}$ and S$_{16}$ 
(Appendix~\ref{app:hp}), with, however, 
zero success rate  and error patterns similar to that for S$_{12}$. 

To further elucidate the impact of excluding larger neighborhoods, we ran QLS using a classical 
solver~\cite{Boppana:92} for such neighborhoods, rather than excluding them. This gave success rates
$\geq$0.6 for all the sequences S$_{4}$--S$_{14}$, and $\geq$0.8 for most of them. 
Furthermore, it permitted us to find the ground states of S$_{15}$--S$_{18}$ (Appendix~\ref{app:hp}), 
with success rates $>$0.3. We conclude that with access to appropriate hardware, our QLS approach 
might be able to find the ground state of even larger lattice proteins. 


It should be stressed that our larger problem sizes are well beyond what can be tackled 
with full-scale QAOA methods; the corresponding quantum circuits are too large 
to simulate. For the S$_{14}$ sequence (Fig.~\ref{fig:successrate_qls}b),
on a $5\times 5$ grid, the qubit count is $NL^2/2=188$.   
By contrast, in our QLS computations for this sequence, the average size of the local neighborhoods 
was 14 qubits when skipping neighborhoods with $>$26 qubits, and 19 qubits when minimizing
classically over large neighborhoods (Fig.~\ref{fig:qls_size}a).

\subsection{Sizes of the local neighborhoods in QLS}\label{sec:results_resources}

By restricting the search to local neighborhoods, the QLS scheme (Sec.~\ref{sec:methods_qls}) 
avoids having to simulate the full qubit system of a given problem instance. The precise size
of the local neighborhoods is stochastic and varies within and between runs. 
\begin{figure}[t]
  \centering
    \includegraphics[width=8cm]{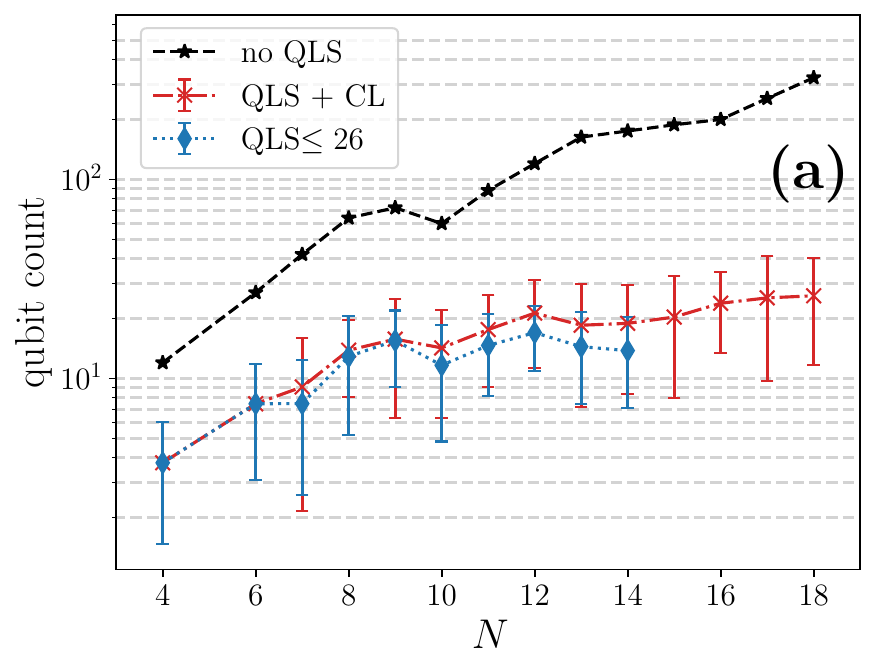}
    \includegraphics[width=8cm]{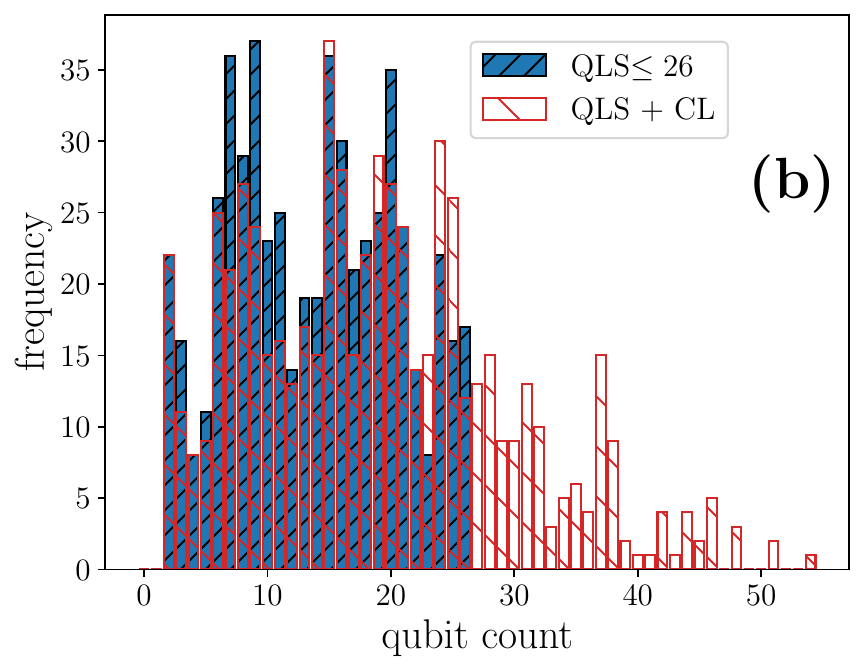}
     \caption{Local neighborhood sizes in our QLS computations (Secs.~\ref{sec:methods_qls}, \ref{sec:results_longer_chains})
     for the sequences  S$_4$--S$_{18}$ (Appendix~\ref{app:hp}).
     (a) Average qubit counts from 10 runs, plotted against the chain length $N$, 
     (i) when skipping neighborhoods with $>$26 qubits (QLS $\leq$26, blue), and 
     (ii) when using all neighborhoods but treating those with $>$26 qubits classically (QLS $+$ CL, red).  
     Black symbols show the size of the full systems. 
     Error bars indicate standard deviations.
     (b) Distribution of neighborhood sizes in the QLS computations for the sequence S$_{14}$.\label{fig:qls_size}}
\end{figure}
Figure~\ref{fig:qls_size}a shows the average size of the local neighborhoods in our QLS computations 
for the sequences S$_4$--S$_{14}$ (Sec.~\ref{sec:results_longer_chains}), when skipping neighborhoods 
with $>$26 qubits. For comparison, it also shows the size of the full qubit systems, and  
the average size of the local neighborhoods when no size limit is imposed. In the latter 
case, we used a classical solver~\cite{Boppana:92} for large neighborhoods. 
For the longest sequences, the average reduction in qubit count is about one order of magnitude upon using the 
local neighborhoods. When comparing the two schemes using local neighborhoods, the classically augmented one 
uses, on average, slightly larger neighborhoods, as expected. However, the differences in average neighborhood size between 
the two schemes are modest. This shows that, while large neighborhoods can have an important role in the 
optimization, most neighborhoods are small even for the longer sequences. 

The conclusion that even when including larger neighborhoods, the majority of 
neighborhoods are small ($\leq$26) is strengthened when comparing the distributions of neighborhood 
sizes for the two schemes. Figure~\ref{fig:qls_size}b shows these two distributions for S$_{14}$. 
From this figure, it can be seen that the classically augmented scheme and the one skipping
larger neighborhoods have strongly overlapping size distributions. However, the former scheme  
has a tail stretching out to larger neighborhoods, even reaching sizes of $>$50 qubits. 
The precise shape of these distributions depends heavily on the pinning parameters.

\section{Summary and discussion}

In this paper, we have described a new QAOA variant for lattice protein folding, which we explored
using classical simulations of the quantum circuits. Our starting point was a QUBO 
formulation of the problem~\cite{Irback:22} in which closed-form penalty energies ensure valid chain 
structures (Sec.~\ref{sec:methods_qubo}). The (positive) quadratic terms in these penalty energies 
can be represented as edges in a conflict graph, with the (qu)bits at its nodes. In this graph representation, 
full-length chain structures correspond to maximum independent sets. It follows that full-length chain structures 
with minimum energy, $\EHP$, can be identified by minimizing the cost function 
$\EMIS=\EHP-\lambda_1\sum_{i,n}b_{i,n}$ over independent sets. While containing a linear bias,
this cost function is free from quadratic penalty terms.   

To implement a QAOA search restricted to independent sets, we adopted a mixer proposed 
for finding maximum independent sets~\cite{Hadfield:19,Tomesh:22,Saleem:23}.  
The resulting QAOA-variant, \qaoamis, was capable of folding short HP sequences ($N=4, 6$)
using circuits with only a few layers. The mixer used relies on multi-qubit controlled gates. These
are currently cumbersome to realize on hardware, but strides are being made toward
efficient implementations~\cite{Silva:22,Zindorf:25}.

To be able to deal with longer chains, we formulated a heuristic iterative scheme, QLS, based on quantum 
local search. This scheme was inspired by previous work on the pure MIS problem~\cite{Tomesh:22}. 
However, in the presence of the protein energy $\EHP$, we found it necessary to take 
a modified approach, which initially focuses on small neighborhoods involving H beads only, and which includes 
a stochastic pinning mechanism to reduce the qubit count. In addition, we skipped neighborhoods with size
above a threshold, set to 26 qubits. Even without this size limit, the restriction to local neighborhoods brings
a significant reduction of the qubit count; for the larger $N$ values studied, the average size of the local neighborhoods        
is about an order of magnitude smaller than the size of the full system (Fig.~\ref{fig:qls_size}a).

We tested this QLS scheme on a set of 10 HP sequences with lengths $4\le N\le 14$, with 
success rates $\ge$0.1 for all sequences. The success rate was lowest for the sequence S$_{12}$,
which had the highest average neighborhood size (Fig.~\ref{fig:qls_size}a). When augmenting the local search 
with a classical solver for neighborhoods with $>$26 qubits, the success rate increased significantly, which
suggests that, with enough qubits, the local search approach can fold long chains. 

Our study shows that creating a conflict graph and optimizing over independent sets 
can be a viable approach to optimization tasks other than the pure MIS problem. The same approach
can be straightforwardly implemented for other optimization problems with similar QUBO formulations, including  
the travelling salesman problem and scheduling problems~\cite{Hopfield:85,Lucas:14,Venturelli:15,Vikstal:20}.

\section*{Acknowledgements}
This work was in part supported by the German Federal Ministry of Education
and Research within the funding program Quantum technologies ---
from basic research to market (contract no. 13N16233).


\begin{appendix}

\section{HP sequences and structures}
\label{app:hp}

\begin{figure}[t]
\centering
	\includegraphics[width=14cm]{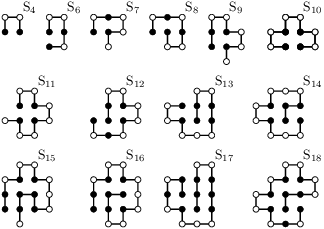}
   \caption{Minimum-energy structures~\cite{Irback:02,Holzgrafe:11}  
   for the HP sequences S$_N$ in Table~\ref{tab:seq}. Filled and open symbols indicate 
   H and P beads, respectively.
   \label{fig:all_chains}}
\end{figure}

From previous work using exhaustive enumerations~\cite{Irback:02,Holzgrafe:11}, 
all $N\le 30$ HP sequences with unique minimum-energy structures are 
known.  For testing \qaoamis, we selected a set of 
such sequences with different lengths, $4\le N\le18$. These sequences, 
denoted by S$_N$, can be found Table~\ref{tab:seq}, and the corresponding 
structures are displayed in Fig.~\ref{fig:all_chains}.    

\begin{table}[H]
  \centering
  \caption{The HP sequences studied, their minimum energies ($\EHPmin$), and 
  the grid sizes used in the full \qaoamis\ (Sec.~\ref{sec:results_short_chains}) and QLS 
  (Sec.~\ref{sec:results_longer_chains}) computations. All sequences have a known 
  unique minimum-energy structure~\cite{Irback:02,Holzgrafe:11}. 
 } 
  \vspace{6pt}
  {\small
  \begin{ruledtabular}
  \begin{tabular}{llcccc}
              &                   &		&  \multicolumn{2}{c}{Grid}\\
     \cline{4-5}
    Name & Sequence & $\EHPmin$ & \qaoamis & QLS\\ 
    \hline
    \vspace{-2pt}
    S$_4$     & HPPH									& $-1$ & $3\times2$ &	$3\times2$\\
    \vspace{-2pt}
    S$_6$     & HPPHPH								& $-2$ & $4\times2$ & 	$3\times3$\\
    \vspace{-2pt}
    S$_7$     & PHPPHPH								& $-2$ && 			$3\times4$\\
    \vspace{-2pt}
    S$_8$     & HPHPHPPH							& $-3$ && 			$4\times4$\\
    \vspace{-2pt}
    S$_9$	   & HHPPHPPHP							& $-3$ && 			$4\times4$\\	
    \vspace{-2pt}
    S$_{10}$ & HPPHPPHPPH              					& $-4$ && 			$3\times4$ \\
    \vspace{-2pt}
    S$_{11}$ & HPPHPPHPPHP             					& $-4$ && 			$4\times4$ \\
    \vspace{-2pt}
    S$_{12}$ & HHPPHPPHPHPH            					& $-5$ && 			$4\times5$ \\
    \vspace{-2pt}
    S$_{13}$ & HHPPHHPPPHPPH            				& $-5$ && 			$5\times5$\\
    \vspace{-2pt}
    S$_{14}$ & HHHPPPHPPHPPPH            			        & $-5$ && 			$5\times5$\\
    \vspace{-2pt}
    S$_{15}$ &  HHPHPPHPPPHHHHP            				& $-6$ && 			$5\times5$\\
    \vspace{-2pt}
    S$_{16}$ &  HHPHPPHPPPHPPHHH      				& $-7$ && 			$5\times5$\\
    \vspace{-2pt}
    S$_{17}$ &  HHHPPHHHPPPHPHPHH					& $-8$ &&				$6\times5$\\
    \vspace{-2pt}
    S$_{18}$ & HHHPPHPPHPHPPHPHPH   				& $-9$ && 			$6\times6$\\
    \vspace{-2pt}
    \vspace{-2pt}
    \vspace{-2pt}
  \end{tabular}	
  \end{ruledtabular}
  }
  \label{tab:seq}
\end{table}

\section{Energy landscapes for single-layer \qaoax\ and \qaoaxy\ circuits\label{app:energy_landscapes}}

\begin{figure}[t]
\centering
	\includegraphics[width=8cm]{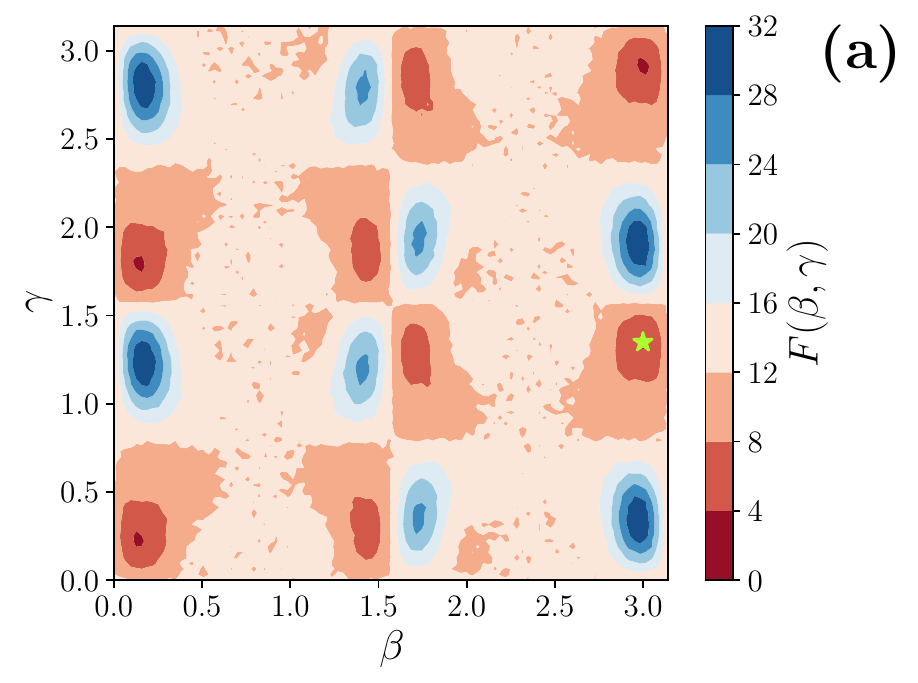}
	\includegraphics[width=8.cm]{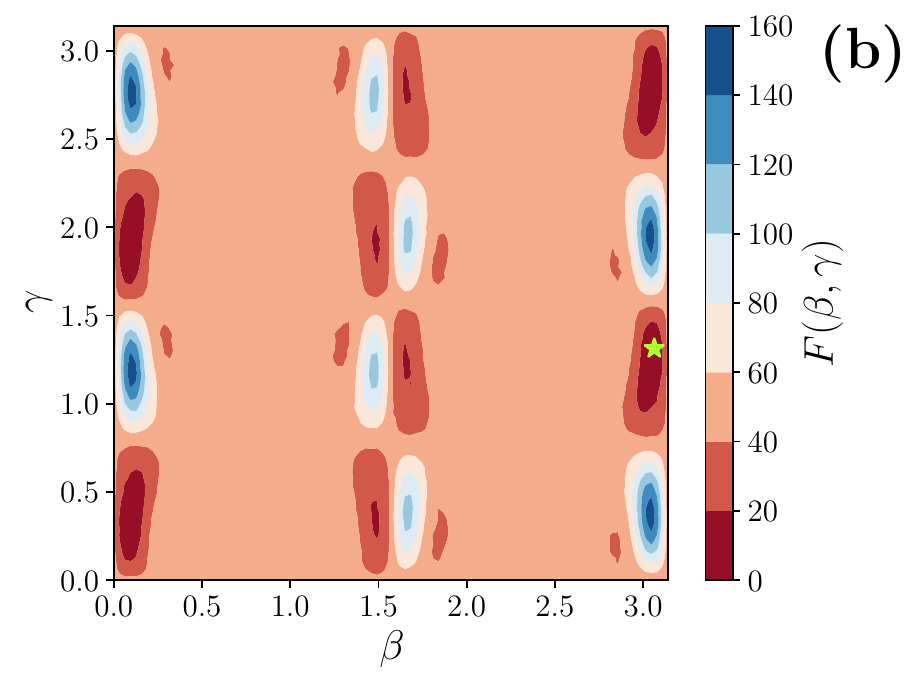} 
   \caption{Energy landscapes $F(\boldsymbol{\theta})$ (Eq.~\ref{eq:F}) for single-layer \qaoax\ circuits $(p=1)$, 
   generated using a $10^2\times10^2$ $\boldsymbol{\theta}$ grid and $10^8$ shots per grid point. 
   Stars indicate starting points for the circuit optimization procedure that gave the   
   best results, shown in Fig.~\ref{fig:small_chains_hitrates}. 
   a) S$_4$ on a $3\times2$ grid. b) S$_6$ on a $4\times2$ grid. 
   \label{fig:landscapes_x}
   }
\end{figure}

\begin{figure}[t]
\centering
	\includegraphics[width=8cm]{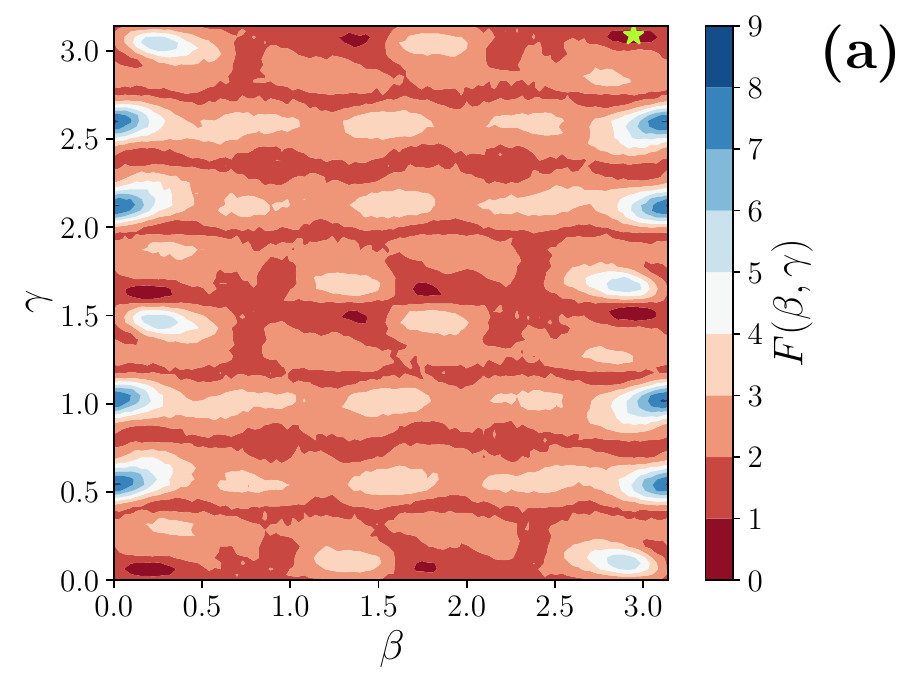}
	\includegraphics[width=8.cm]{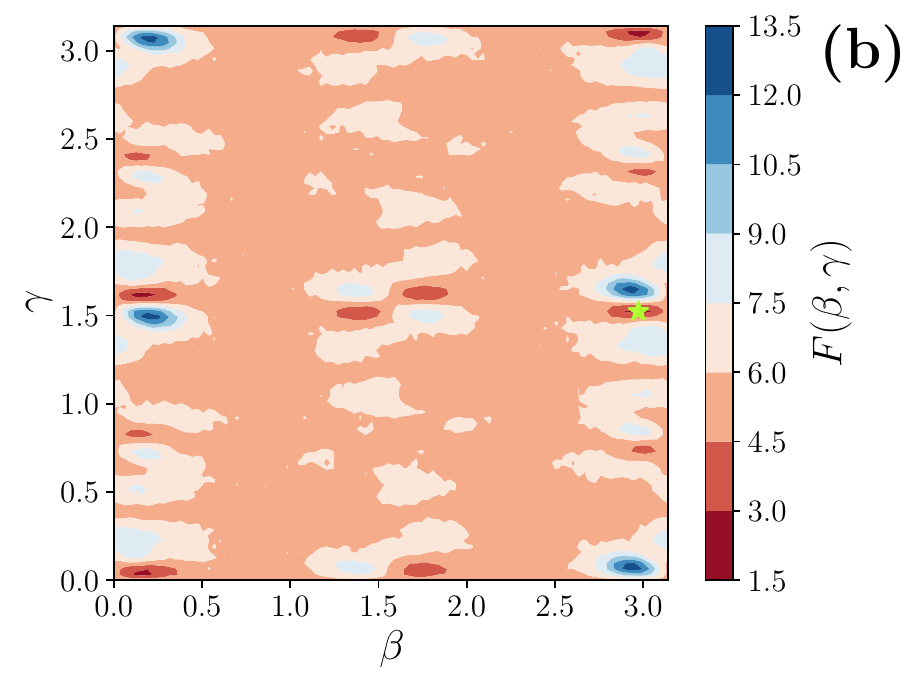} 
   \caption{ 
   Energy landscapes $F(\boldsymbol{\theta})$ (Eq.~\ref{eq:F}) for single-layer \qaoaxy\ circuits $(p=1)$, 
   generated using a $10^2\times10^2$ $\boldsymbol{\theta}$ grid and $10^8$ shots per grid point. 
   Stars indicate starting points for the circuit optimization procedure that gave the   
   best results, shown in Fig.~\ref{fig:small_chains_hitrates}. 
   a) S$_4$ on a $3\times2$ grid. b) S$_6$ on a $4\times2$ grid. 
   \label{fig:landscapes_xy}
   }
\end{figure}

For \qaoax\ and \qaoaxy\ circuits with a single layer ($p=1$), there are only two variational parameters, $\boldsymbol{\theta}=(\beta,\gamma)$,
which makes it possible to do a grid-based scan of the energy landscape $F(\boldsymbol{\theta})$ (Eq.~\ref{eq:F}). Using a $10^2\times10^2$  
$\boldsymbol{\theta}$ grid, we performed such scans for the S$_4$ and S$_6$ systems discussed in Sec.~\ref{sec:results_short_chains}.  
The resulting energy landscapes are shown in Figs.~\ref{fig:landscapes_x} and \ref{fig:landscapes_xy}. 
Each of the generated energy landscapes exhibits a limited number of hills and cavities, and is otherwise quite flat. 

For a given QAOA variant and a given problem instance, we picked one point from each of the four deepest cavities 
in the energy landscape as starting point for a deterministic circuit optimization procedure. Here, we first optimized 
the $p=1$ circuit, and then iteratively optimized circuits with increasingly large $p$, up to $p=20$, 
using parameter transfer through interpolation~\cite{Zhou:20}. In all cases, one of the four starting points led to
consistently higher success probabilities for the different values of $p$. The success probabilities shown in Fig.~\ref{fig:small_chains_hitrates} 
come from the runs with best starting points. These starting points are indicated by a star in Figs.~\ref{fig:landscapes_x} and \ref{fig:landscapes_xy}.
 
\end{appendix}
\clearpage

\bibliography{references}
\end{document}